\documentclass[10pt]{iopart}
\usepackage[OT1,T1]{fontenc}
\usepackage[latin1]{inputenc}
\usepackage{iopams} 
\usepackage{bbm} 
\usepackage{undertilde} 
\newcommand{\di}{\mathrm{d}} 
\newcommand{\ou}[3]{{#1}{}^{#2}{}_{#3}} 
\newcommand{\uo}[3]{{#1}{}_{#2}{}^{#3}} 
\newcommand{\I}{\mathrm{i}} 
 
\newcommand{\ellp}{\ell_{\mathrm{P}}} 
\newcommand{\odens}{\widetilde{\eta}} 
\newcommand{\udens}{\utilde{\eta}} 
\newcommand{\qq}[1]{``#1''} 
  
\newcommand{\klamm}{\;\right\}} 
\usepackage{hyperref}

\begin{document}

\title[Simplified Hamiltonian constraint for a particular value of $\beta$]{Simplified Hamiltonian constraint for a particular value of the Barbero--Immirzi parameter}
\author{Herbert Balasin${}^1$ and Wolfgang M. Wieland${}^2$}
\address{Institute for Theoretical Physics,
   Vienna University of Technology\\
   Wiedner Hauptstra\ss e 8-10/136, 1040 Wien, Austria}
\eads{${}^1$\mailto{hbalasin@tph.itp.tuwien.ac.at}, \mailto{${}^2$wieland@tph.itp.tuwien.ac.at}}
\begin{abstract}
  Investigating iterated Poisson brackets involving the volume functional 
together with the Euclidean Hamiltonian constraint for unit lapse we show 
that within the Ashtekar formulation of General Relativity, 
the value $\beta=1/\sqrt{3}$ of the Barbero--Immirzi parameter 
corresponds to a considerable simple and compact form of the Hamiltonian 
constraint. 
This value might yield a new starting point for the quantization of the 
Hamiltonian constraint of General Relativity.
\end{abstract}

\section{Introduction}
Loop quantum gravity (LQG for short, in \cite{ashtekar, rovelli, thiemann, status} introductions can be found) is one of the most promising candidates for a quantum theory of the gravitational field. It represents a non-perturbative,
mathematically rigorous quantization of General Relativity (GR) in the canonical framework.
The corresponding phase space variables are the $\mathfrak{su}(2)$-valued Ashtekar connection $\ou{A}{i}{a}$
together with its momentum conjugate $\uo{E}{i}{a}$ commonly called densitized triad or (Yang-Mills) electric field for short. 

The crucial step for quantizing GR in the Ashtekar framework arises from the natural coupling of the
connection to oriented curves $\gamma$ via the so-called holonomy map $h_{\gamma}[A]=\mathcal{P}\mathrm{exp}({-\int_{\gamma}A})$
i.e. the path-ordered exponential of the integrated Ashtekar connection $A$. This in turn allows for a rigorous construction of a Hilbert space (more precisely a Gel'fand triplet) as an $L^{2}$-space relative to the Ashtekar--Lewandowski measure.
One of the great surprises and key results of the theory comes from the fact that certain geometrical
quantities, most notably the volume of a region and the area of a surface, can be turned 
into well-defined operators on this Hilbert space, which are self-adjoint and posses a discrete spectrum, thereby exhibiting
the discrete structure of space at the fundamental Planck scale. 

In order to regain the dynamics of GR at the quantum level one has to turn the classical constraints into operators and find their corresponding
solution space. Due to the background independent construction this can be done with remarkable ease for the so-called Gau\ss\; and vector (or diffeomorphism) constraints,
which represent the kinematical symmetries of the theory. The main stumbling block that remained was the Hamiltonian (or time-evolution)
constraint, which turned out to be quite difficult to quantize (i.e. define an operator version). Once again the interplay
between kinematical symmetries and the existence of a well-defined quantization of the volume functional, together with the ingenious insight
of Thiemann allows the implementation of the Hamiltonian constraint as a well-defined operator. 
Thiemann's discovery rests on the fact that the presence of an inverse volume form in the (Euclidean part of the) Hamiltonian constraint compensating for the double-density behavior of two electric fields, instead of being disastrous could actually be turned into an advantage. This is achieved by replacing all three terms (i.e. the two electric fields together with the inverse volume form) by a single Poisson bracket with the volume functional. Actually the same techniques
apply to the so-called kinetic part, which however rests on the existence of the Euclidean part as a first step. 

The present proposal tries to investigate the structure of the full Hamiltonian constraint under the light of the above insight. We will show that it is actually possible to rewrite the Hamiltonian constraint in a very compact form, namely as twice iterated Poisson bracket of
the Euclidean Hamiltonian constraint and the \qq{smeared} volume functional. This draws on the simple observation that the lapse function allows
redistribution from the Hamiltonian to the volume. It has to be pointed out that the proposal works only for the definite value $\beta=1/\sqrt{3}$ of the so-called
Barbero--Immirzi parameter, which measures the relative strength between the Ashtekar and the spin-connection.

In the remaining part of this section we will briefly define the notation that will be used subsequently. Later in section \ref{sectiontwo}, the main part of this paper  we present a simplification of the Hamiltonian constraint, provided the Barbero--Immirzi Parameter takes the special value $\beta=1/\sqrt{3}$. In the last section with a summary and discussion we conclude this paper. 
\paragraph{Notation} 
Let us first define the Planck length according to $\ellp:=\sqrt{8\pi\hbar G/c^3}\approx 8.1\cdot 10^{-35}\mathrm{m}$. In the following we will make heavily use of abstract index notation, small roman indices from the beginning of our alphabet ($a,b,c,\dots$) refer to abstract indices \cite{wald} in the tangent space $T\Sigma$ of the 3-dimensional $t=\mathrm{const.}$ Cauchy surface $\Sigma$. Indices from the middle of the alphabet ($i,j,k,\dots$) refer to a decomposition of the vector space $\mathfrak{su}(2)$ with respect to the standard basis $\tau_i:=\frac{1}{2\I}\sigma_i$ (and $\sigma_i$ are the Pauli matrices). Small greek indices correspond to some local coordinate functions $x^\mu$. Using abstract index notation the (co-)basis vectors $\partial/\partial x^\mu\in T\Sigma$ (and $\di x^\mu\in T^\ast\Sigma$) can be written according to $\partial^a_\mu\in T\Sigma^a$ (and $\di x^\mu_a\in T\Sigma_a$  respectively). 

With respect to tensor densities we use the following notation. Given the volume element $\omega_h=1/3!\,\varepsilon_{\mu\nu\rho}\di x^\mu\wedge\di x^\mu\wedge\di x^\rho$ ($\varepsilon_{\mu\nu\rho}$ are the components of the spatial Levi-Civita tensor with respect to the coordinate basis chosen) any tensor density can be decomposed in a tensor times the density $\omega_h$. Furthermore on $\Sigma$ there is a natural metric independent three-vector density of weight one:
\begin{equation}
 \odens^{abc}:=\di x^\mu\wedge\di x^\nu\wedge\di x^\rho\partial^a_\mu\partial^b_\nu\partial^c_\rho
\end{equation}
Its inverse---a three-form density of weight minus one---we denote by $\udens_{abc}$, it fulfils:
\begin{equation}
 \odens^{a_1a_2a_3}\udens_{b_1b_2b_3}=\sum_{\pi\in S_3}\mathrm{sign}(\pi)\,\delta^{a_{\pi(1)}}_{b_1}\delta^{a_{\pi(2)}}_{b_2}\delta^{a_{\pi(3)}}_{b_3}
\end{equation}
And $S_3$ is the group of permutations of three symbols. 
Furthermore $e^i$ (and $\ou{e}{i}{a}$ in the case of abstract index notation) is the co-triad or soldering form, from which the spatial metric is reconstructed according to $h_{ab}=\delta_{ij}\ou{e}{i}{a}\ou{e}{j}{b}$. The inverse we always denote by $\uo{e}{i}{a}$ and call it triad. From the co-triad an \emph{oriented} volume element is built according to:
\begin{equation}
 \omega_E:=\frac{1}{3!}\epsilon_{ijk}e^i\wedge e^j\wedge e^k=e^1\wedge e^2\wedge e^3=\varepsilon\omega_h
\end{equation}
Here $\epsilon_{ijk}$ denotes the totally antisymmetric symbol in internal space---fixed once and for all to the values $\epsilon_{123}=1$, $\epsilon_{132}=-1$ and so on. Furthermore $\varepsilon$ is the orientation of the triple $(e^1, e^2, e^3)$ which can in General Relativity always be fixed to either $\pm 1$ globally. The spatial Levi-Civita covariant derivative we denote by $D_a$, acting both on internal $\mathfrak{su}(2)$ and tangent indices it annihilates the triad $D_a\uo{e}{i}{b}=0$ and its inverse too.
To abbreviate iterated Poisson brackets we define for $n\geq0$ that $\{A,B\}^{n+1}=\{\{A,B\}^n,B\}$, together with $\{A,B\}^0=A$.
\section{Poisson brackets containing volume and Euclidean Hamiltonian}\label{sectiontwo}
\subsection{Ashtekar variables---an elementary overview}
The Ashtekar variables are related to geometrodynamics in the ADM-formulation \cite{adm, wald} via:
\begin{equation}
 \uo{E}{i}{a}=\omega_h\uo{e}{i}{a}, \qquad\ou{A}{i}{a}=\ou{\Gamma}{i}{a}[E]+\beta\varepsilon\ou{K}{i}{a}\label{ashvar}
\end{equation}
Here $K_{ab}$ in the form of $\ou{K}{i}{a}=e^{ib}K_{ba}$ is the extrinsic curvature tensor (the second fundamental form of $\Sigma$), the quantity $\ou{\Gamma}{i}{a}[E]$ denotes the three dimensional spin connection 1-form, decomposed according to $\ou{\Gamma}{i}{ja}=\ou{[T_k]}{i}{j}\ou{\Gamma}{k}{a}$ into the generators $\ou{[T_i]}{j}{k}=\ou{\epsilon}{j}{ik}$ of $SO(3)$. Its dependence on the densitized triad is determined by Cartan's first structure equation (i.e. the vanishing of the torsion 2-form $D\wedge e^i=\di\wedge e^i+\ou{\epsilon}{i}{jk}\Gamma^j\wedge e^k=0$). Forming a canonical pair the only non vanishing Poisson bracket of the elementary variables reads:
\begin{equation}
 \big\{\uo{E}{i}{a}(p),\ou{A}{j}{b}(q)\big\}=\gamma\delta^j_i\delta^a_b\delta^{(3)}(p,q),\quad \gamma:=\frac{\beta\ellp^2}{\hbar}\label{sympstruct} 
\end{equation}
Furthermore $\beta$ (the Barbero--Immirzi parameter) is a new dimensionless positive constant of nature. Introducing this parameter only effects the quantization of GR, the classical theory is completely insensitive to its value. From the black hole entropy calculation in the LQG-framework \cite{bhentashkrasn} together with the requirement to precisely reproduce the Bekenstein--Hawking formula $S_{\mathrm{BH}}=\frac{1}{4}\frac{k_{\mathrm{B}}c^3}{\hbar G}A_H$ the authors of \cite{parametercorichi} propose this parameter to be fixed to the value:
\begin{equation}
 \beta\approx 0.2374 \label{bivalue}
\end{equation}
By the very same argument one finds in \cite{parametermeissner, thiemann} a slightly different value of $\beta\approx 0.2375$ instead, which holds for all stationary black holes. 

One might wonder now about the orientation factor in our definition of the Ashtekar variables \eref{ashvar}. It was introduced to get rid of additional constraints the Ashtekar variables would have to fulfil otherwise. Consider for example the pair ${E^\prime_i}{}^{a}:=\omega_E\uo{e}{i}{a}$ and $\ou{A^\prime}{i}{a}:=\ou{\Gamma}{i}{a}+\beta\ou{K}{i}{a}$ instead. Let $\mathcal{P}$ be the parity transformation in internal space (i.e. the map $\mathcal{P}:e^i\mapsto-e^i$). If $\ou{\mathcal{P}A^\prime}{i}{a}$ is the parity transformed connection, the following additional constraints would have now to be fulfiled:
\numparts
\begin{eqnarray}
   \udens_{abc}\epsilon^{ijk}{E^\prime_i}{}^{a}{E^\prime_j}{}^{b}{E^\prime_k}{}^{c} \stackrel{!}{>}  0 \\
   \frac{1}{2}\big(\ou{A^\prime}{i}{a}+\ou{\mathcal{P}A^\prime}{i}{a}\big) \stackrel{!}{=} \ou{\Gamma}{i}{a}[E^\prime] 
\end{eqnarray}
\endnumparts
These addiditonal  constraints are absent in \eref{ashvar}, and could not straight forwardly be implemented in the quantum theory. One can find a short discussion of the same subject in appendix A of \cite{bianchiiilqg}. 

Furthermore the Ashtekar variables in the form of \eref{ashvar} most naturally emerge from the Hamiltonian formulation of the 1995 discovered \qq{Holst action} introduced in \cite{holst}, which takes the following form:
\begin{equation}
 S_\mathrm{H}[\eta,\omega]=\frac{\hbar}{4\ellp^2} 
        \int_M\Big(\epsilon_{IJKL}\eta^I\wedge\eta^J\wedge\Omega^{KL}[\omega]
        - \frac{2\varepsilon}{\beta}\eta_I\wedge\eta_J\wedge\Omega^{IJ}\Big)
\end{equation} 
Here $\eta$ denotes the co-tetrad (the four dimensional extension of $e$), $\omega$ is the $\mathfrak{so}(1,3)$ valued spin connection, $\Omega$ is its curvature, and $\epsilon_{IJKL}$ is the four dimensional antisymmmetrising operator in internal space (fixed to the values $\epsilon_{0123}=1$, $\epsilon_{1023}=-1$ etc). Roughly speaking the Lagrangian associated differs from the Einstein--Hilbert Lagrangian $\mathcal{L}_{\mathrm{EH}}[g]=\di^4x\sqrt{-g}R$ by the orientation $\varepsilon$ of the co-tetrad $\mathcal{L}_{\mathrm{EH}}[g[\eta]]=\varepsilon\mathcal{L}_{\mathrm{H}}[\eta,\omega[\eta]]$. Here $\omega[\eta]$ has to be understood as the four dimensional Levi-Civita spin connection, functionally depending on the tetrad, and $g[\eta]=\eta_{IJ}\eta^I\otimes\eta^J$ is the space time metric associated. 
Furthermore the orientation factor introduced in the second term of the above equation absent in the original definition of the Holst action is needed to recover the Ashtekar variables defined as in \eref{ashvar}, and ensures all terms to transform equally under an internal reflection. 
By this definition parity is neither a symmetry of the symplectic structure \eref{sympstruct} nor of the Hamiltonian constraint being defined by (\ref{iii}) and equations \eref{euham}, \eref{kinham}, and \eref{cosm}.

The phase space built from all pairs $(\uo{E}{i}{a},\ou{A}{i}{a})$ contains unphysical degrees of freedom, configurations being realised in nature are selected by the vanishing of three types of constraints, Gau\ss\,, vector and Hamiltonian constraint respectively. The smeared versions of the constraints take the following general form:
\begin{enumerate}
 \item Gau\ss\, constraint: $G_i[\Lambda^i]:=\int_\Sigma \Lambda^i\mathcal{D}_a\uo{E}{i}{a}\stackrel{!}{=}0$, for all $\Lambda^i$.
 \item Vector constraint: $H_a[V^a]:=\int_\Sigma V^a\ou{F}{j}{ab}\uo{E}{j}{b}\stackrel{!}{=}0$, for all $V^a$.
 \item Hamiltonian constraint: $H^{\beta}[N]:=H_{\mathrm{E}}[N]+(\beta^2+1)T[N]\stackrel{!}{=}0$, for all $N$.\label{iii}
\end{enumerate}
Here $\Lambda^i:\Sigma\rightarrow\mathfrak{su}(2)$, $V^a:\Sigma\rightarrow T\Sigma^a$ and $N:\Sigma\rightarrow\mathbb{R}$ are smooth functions of compact support, $\mathcal{D}_a=\partial_a+[A_a,\cdot]$ is the covariant derivative associated to the Ashtekar connection, and $\ou{F}{j}{ab}$ is its field-strength (or curvature):
\begin{equation}
  F=\di\wedge A+\frac{1}{2}\big[A,A\big]\Leftrightarrow\ou{F}{i}{ab}=\partial_a\ou{A}{i}{b}-\partial_b\ou{A}{i}{a}+\ou{\epsilon}{i}{jk}\ou{A}{j}{a}\ou{A}{k}{b}
\end{equation}
The Hamiltonian constraint splits into three terms, the first two of them are called Euclidean and \qq{kinetic} Hamiltonian respectively, they take the following form:
\numparts
\begin{eqnarray}
 \textnormal{Euclidean part:}  \quad & H_{\mathrm{E}}[N]=\frac{1}{2}\int_\Sigma N\ou{F}{j}{ab}\uo{\odens}{j}{ab}\label{euham}\\
 \textnormal{Kinetic part:}    \quad & T[N]=-\frac{1}{2}\int_\Sigma N \ou{\epsilon}{j}{im}\ou{K}{i}{a}\ou{K}{m}{b}\uo{\odens}{j}{ab}\label{kinham}
\end{eqnarray}
\endnumparts
Simplyfying our notation we have introduced the $\mathfrak{su}(2)$ valued bi-vector density $\uo{\odens}{j}{ab}=(\omega_E)^{-1}\uo{\epsilon}{j}{lm}\uo{E}{l}{a}\uo{E}{m}{b}$. 
Using the \qq{determinant} formula $\omega_E\epsilon^{ijk}\uo{e}{i}{a}\uo{e}{j}{b}\uo{e}{k}{c}=\odens^{abc}$ the Euclidean Hamiltonian can be rewritten according to the following equation:
\begin{equation} 
 H_{\mathrm{E}}[N]=\frac{1}{2}\int_\Sigma N \odens^{abc}\ou{F}{j}{ab}e_{jc}
\end{equation}
It is precisely this form that was employed by Thiemann as starting point that allowed the quantisation to proceed \cite{qsd}.
\subsection{Variations on the phase space}
This subsection will be devoted to the calculation of exterior derivatives $\mathbbm{d}F$ (i.e. \qq{variations}) of some (assumed to be differentiable) functionals on the phase space of the Ashtekar variables. Let us first study what may be called \qq{smeared} volume of space, i.e. the quantity
\begin{equation}
 \boldsymbol{V}[\varepsilon N]:=\int_\Sigma N\omega_E=\int_\Sigma N\varepsilon\omega_h\label{cosm}
\end{equation}
The metrical volume three form $\omega_h$ which in terms of the Ashtekar variables can be rewritten as the square root of the \qq{determinant} of the densitized triad
\begin{equation}
 \omega_h=\sqrt{\Big|\frac{1}{3!}\udens_{abc}\epsilon^{ijk}\uo{E}{i}{a}\uo{E}{j}{b}\uo{E}{k}{c}\Big|},
\end{equation}
which yields the functional differential of the smeared volume of space
\begin{equation}
 \mathbbm{d}\boldsymbol{V}[N]=\frac{1}{2}\int_\Sigma N \ou{e}{i}{a}\mathbbm{d}\uo{E}{i}{a}.
\end{equation}
Here the variation of the orientation of the triad has been neglected. In the classical theory this factor can however always be fixed to $\pm1$ globally, and we can thus neglect all variations and derivatives associated. Otherwise computing Poisson brackets we could encounter singular (distributional) expressions of the form of $\int_\Sigma\partial_a\varepsilon[\dots]$ leading to generically non vanishing surface terms. By the same argument we find:
\begin{equation}
 \mathbbm{d}\omega_E=\frac{1}{2}\varepsilon\ou{e}{i}{a}\mathbbm{d}\uo{E}{i}{a}\label{vediff}
\end{equation}
Following Thiemann we define the smeared version of the integrated trace of the extrinsic curvature tensor:
\begin{equation}
 K[N]:=\int_\Sigma\beta\varepsilon N\ou{K}{i}{a}\uo{E}{i}{a}=\int_\Sigma N\big(\ou{A}{i}{a}-\ou{\Gamma}{i}{a}[E]\big)\uo{E}{i}{a}
\end{equation}
Computing the functional differential we recall the definition $\di\wedge e^i+\ou{\epsilon}{i}{jk}\Gamma^j\wedge e^k=0$ of the Ricci-rotation coefficients $\ou{\Gamma}{i}{a}$ and find:
\begin{equation}
 D\wedge \mathbbm{d}e^i+\ou{\epsilon}{i}{jk}\mathbbm{d}\Gamma^j\wedge e^k=0
\end{equation}
From this equation we obtain after performing a partial integration (remember $N$ to be a scalar test function) and using the fact that the covariant spatial Levi-Civita derivative acting on all indices annihilates the (co-)triad (i.e. $D_a\uo{e}{i}{b}=0$ and $D_a\ou{e}{i}{b}=0$) together with the identity $\mathbbm{d}(e_{lb}\uo{e}{m}{b})=0$ that:
\begin{equation}
 \left.\fl\eqalign{\int_\Sigma N&\mathbbm{d}\ou{\Gamma}{i}{a}\uo{E}{i}{a}
     =   \frac{1}{2}\int_\Sigma N\varepsilon\epsilon_{mil}\mathbbm{d}\ou{\Gamma}{i}{a}\ou{e}{l}{b}\ou{e}{m}{c}\odens^{abc}
         =-\frac{1}{2}\int_\Sigma N\varepsilon D_a\mathbbm{d}e_{mb}\ou{e}{m}{c}\odens^{abc}=\\
   & =  \frac{1}{2}\int_\Sigma\varepsilon\partial_aN\mathbbm{d}e_{mb}\ou{e}{m}{c}\odens^{abc}
         =\frac{1}{2}\int_\Sigma\omega_h\partial_a N\mathbbm{d}e_{mb}\epsilon^{ijm}\uo{e}{i}{a}\uo{e}{j}{b}=\\
   & =  -\frac{1}{2}\int_\Sigma\omega_h\partial_a Ne_{mb}\epsilon^{ijm}\uo{e}{i}{a}\mathbbm{d}\uo{e}{j}{b}
         =\frac{1}{2}\int_\Sigma\epsilon^{imj}\uo{e}{i}{a}\partial_a Ne_{mb}\mathbbm{d}\uo{E}{j}{b}}\;\klamm
\end{equation}
To arrive at the latter one needs the \qq{determinant} formula as well the identity $\mathbbm{d}(e_{lb}\uo{e}{m}{b})=0$. We now know the functional differential of $K[N]$ to be:
\begin{equation}
  \mathbbm{d}K[N] = \int_\Sigma N\mathbbm{d}\ou{A}{i}{a}\uo{E}{i}{a}+N\beta\varepsilon\ou{K}{i}{a}\mathbbm{d}\uo{E}{i}{a}
                   -\frac{1}{2}\epsilon^{lmi}\uo{e}{l}{b}\partial_bNe_{ma}\mathbbm{d}\uo{E}{i}{a}\;\;
\end{equation}
Finally we derive the functional differential of the Euclidean Hamiltonian, taking into account the variation of the curvature
\begin{equation}
 \mathbbm{d}F=\mathcal{D}\wedge \mathbbm{d}A\Leftrightarrow\mathbbm{d}\ou{F}{i}{ab}=\mathcal{D}_a\mathbbm{d}\ou{A}{i}{b}-\mathcal{D}_b\mathbbm{d}\ou{A}{i}{a}\label{fdiff},
\end{equation}
and using equations \eref{vediff} and \eref{fdiff} together with the identity $\mathcal{D}\wedge e^i=(\mathcal{D}-D)\wedge e^i=\varepsilon\beta [K,e]^i$, we find after a partial integration
\begin{equation}
 \fl\eqalign{\mathbbm{d}H_{\mathrm{E}}[N]
   & = \int_\Sigma \Big[ -\frac{N}{4}(\omega_h)^{-1}\ou{e}{i}{a}\mathbbm{d}\uo{E}{i}{a}\odens^{bcd}\ou{F}{j}{bc}e_{jd}+
                                 N(\omega_E)^{-1}\uo{\epsilon}{j}{im}\ou{F}{j}{ab}\mathbbm{d}\uo{E}{i}{a}\uo{E}{m}{b}+\\
   &   \quad-N\odens^{bac}\mathbbm{d}\ou{A}{i}{a}\epsilon_{ilm}\beta\varepsilon\ou{K}{l}{b}\ou{e}{m}{c}
                                 -\partial_bN\odens^{bac}\mathbbm{d}\ou{A}{i}{a}e_{ic}\Big]}
\end{equation}
\subsection{A series of Poisson brackets} Using the preparations of the last section we are now ready to compute the second iterated Poisson bracket between volume and Euclidean Hamiltonian, more precisely the quantity $\{\boldsymbol{V},H_{\mathrm{E}}\}^2$. Afterwards using this result we will reconstruct the full Hamiltonian constraint for $\beta=1/\sqrt{3}$ obtaining a surprisingly simple expression. First let us however revisite the following expression, which can of course already be found in \cite{qsd} as well: 
\begin{equation}
 \eqalign{\big\{\boldsymbol{V}[N],H_{\mathrm{E}}[M]\big\} 
   & = -\frac{\beta\gamma}{2}\int_\Sigma MN\varepsilon\odens^{bac}\ou{e}{i}{a}\ou{e}{m}{c}\epsilon_{ilm}\ou{K}{l}{b}\\
   & = \gamma K[MN\varepsilon]}\label{VHE}
\end{equation}
To us the following expression being the second iterated Poisson bracket between smeared volume of space and Euclidean Hamiltonian in the form of  $\{\boldsymbol{V}[N],H_{\mathrm{E}}[M]\}^2$ is of much greater interest:
\begin{equation}
 \fl\left.\eqalign{&\big\{K[N],H_{\mathrm{E}}[M]\big\} 
       = \gamma\int_\Sigma\Big[N\beta\varepsilon\ou{K}{i}{a}
              -\frac{1}{2}\partial_bN\epsilon^{lmi}\uo{e}{l}{b}e_{ma}\Big]\cdot\\
      & \quad \cdot\Big[-M\odens^{dac}\epsilon_{irs}\beta\varepsilon\ou{K}{r}{d}\ou{e}{s}{c}
              -\partial_dM\odens^{dac}e_{ic}\Big]+\\
      & \quad -\gamma\int_\Sigma MN\uo{E}{i}{a}\Big[-\frac{1}{4}(\omega_h)^{-1}\ou{e}{i}{a}\odens^{bcd}\ou{F}{j}{bc}e_{jd}
              +(\omega_E)^{-1}\uo{\epsilon}{j}{im}\ou{F}{j}{ab}\uo{E}{m}{b}\Big]=\\
      & = \gamma\int_\Sigma \Big[MN\beta^2\odens^{adc}\epsilon_{irs}\ou{K}{i}{a}\ou{K}{r}{d}\ou{e}{s}{c}
              +N\partial_dM\beta\varepsilon\odens^{dca}K_{ca}+\\
      & \quad +\frac{1}{2}M\partial_bN\beta\omega_h\epsilon^{ilm}\epsilon_{irs}\uo{\epsilon}{jm}{s}\uo{e}{l}{b}\ou{K}{r}{d}e^{jd}
              +\frac{1}{2}\partial_dM\partial_bN\epsilon^{ilm}\odens^{dac}e_{ic}\uo{e}{l}{b}e_{ma}+\\
      & \quad +\frac{3}{4}MN\odens^{bcd}\ou{F}{j}{bc}e_{jd}-MN(\omega_E)^{-1}\uo{\epsilon}{j}{im}\ou{F}{j}{ab}\uo{E}{i}{a}\uo{E}{m}{b}\Big]=\\
      & = \gamma\int_\Sigma\Big[MN\beta^2(\omega_E)^{-1}\uo{\epsilon}{j}{lm}\ou{\epsilon}{j}{ir}\ou{K}{i}{a}\ou{K}{r}{b}\uo{E}{l}{a}\uo{E}{m}{b}
              +\partial_aMN\varepsilon\uo{e}{i}{a}\mathcal{D}_bE^{ib}+\\
      & \quad +\frac{1}{2}M\partial_b N\varepsilon\uo{e}{l}{b}\mathcal{D}_aE^{la}
              +h^{ab}\partial_aM\partial_bN\omega_E-\frac{1}{4}MN(\omega_E)^{-1}\uo{\epsilon}{j}{im}\ou{F}{j}{ab}\uo{E}{i}{a}\uo{E}{m}{b}\Big]}\klamm
\end{equation}
The latter can be written more compactly according to:
\begin{equation}
  \eqalign{\big\{K[N],&H_{\mathrm{E}}[M]\big\} = -2\gamma\beta^2T[MN]-\frac{\gamma}{2}H_{\mathrm{E}}[MN]+\\
         & + \frac{1}{2}G_i\big[\varepsilon e^{ia}(2\partial_aMN+M\partial_aN)\big]+\gamma\boldsymbol{V}[\varepsilon h^{ab}\partial_aM\partial_bN]}\label{VHEE}
\end{equation}
The last step made use of the Gau\ss\; constraint which restricts $K_{ab}$ to be symmetric:
\numparts
\begin{eqnarray}
  & & \mathcal{D}_a\uo{E}{i}{a}=(\mathcal{D}_a-D_a)\uo{E}{i}{a}=\beta\omega_E\uo{\epsilon}{il}{m}\ou{K}{l}{a}\uo{e}{m}{a}\\
  & & K_{ab}-K_{ba}=\frac{1}{\beta}\mathcal{D}_d\uo{E}{i}{d}e^{ic}\udens_{abc}
\end{eqnarray}
\endnumparts
From \eref{VHE} together with \eref{VHEE} by taking $M=1$ we immediately find that:
\begin{equation}
   \eqalign{\!\!\!\Big\{\big\{
      & \boldsymbol{V}[N],H_{\mathrm{E}}[1]\big\},H_{\mathrm{E}}[1]\Big\}=
              -\frac{\gamma^2}{4}\int_\Sigma N(\omega_h)^{-1}\Big[\uo{\epsilon}{j}{mn}\ou{F}{j}{ab}\uo{E}{m}{a}\uo{E}{n}{b}\\
      & -4\beta^2\ou{\epsilon}{j}{rs}\uo{\epsilon}{j}{mn}\ou{K}{r}{a}\ou{K}{s}{b}\uo{E}{m}{a}\uo{E}{n}{b}\Big]
              +\frac{\gamma^2}{2}\int_\Sigma e^{ia}\partial_aN\mathcal{D}_b\uo{E}{i}{b}}
\end{equation}
Observe now that if
\begin{equation}
   4\beta^2={\beta^2+1}\Leftrightarrow \beta=\pm\frac{1}{\sqrt{3}}\approx\pm 0.5774,
\end{equation}
the above equation reproduces the full Hamiltonian constraint. To be more precisely we have found the following identity:
\begin{equation}
   \textnormal{for}\,\beta=\frac{1}{\sqrt{3}}:\quad H^{\beta}[\varepsilon N]=
            -\frac{2}{\gamma^2}\big\{\boldsymbol{V}[N],H_{\mathrm{E}}[1]\big\}^2
            +G_i\big[e^{ia}\partial_aN\big]\label{Hpropsd}
\end{equation}
Observe $H^{\beta}[\varepsilon N]$ to be insensitive to the orientation of the triad. This follows from the fact that with respect to an internal reflection (i.e. a parity transformation) the densitized triad $\uo{E}{i}{a}$ is changed by an overall minus, but $\ou{F}{j}{ab}$---an internal pseudovector---remains unchanged. 
In the case of $\beta=1/\sqrt{3}$ we have recovered the Hamiltonian constraint associated to the Einstein--Hilbert action---which in contrast to the Hamiltonian constraint derived from the Holst action does not know anything of the orientation of the triad at all.
\section{Conclusion}
Let $\mathcal{K}$ denote the kinematical Hilbert space of LQG, that is the completion of the vector space (with respect to the norm induced by the natural inner product thereof) spanned by all $SU(2)$ gauge invariant spin network states, and let $\Psi$ be some vector therein. On $\mathcal{K}$ there is a quantization of both volume (see \cite{ashvolume} and \cite{rovelli} for two slightly different approaches to the quantization of volume) and a quantization $\widehat{H_{\mathrm{E}}}[N]$ of the Euclidean part of the Hamiltonian constraint for generic lapse $N$ (see for instance \cite{qsd}). In the case the Barbero--Immirzi parameter takes the special value $\beta=1/\sqrt{3}$ (which differs from the proposals in \cite{parameterrovelli, parametercorichi, parametermeissner} considerably) we would thus be allowed to propose the following quantization of the full Hamiltonian constraint of General Relativity:
\begin{equation}
        \textnormal{for}\,\beta=\frac{1}{\sqrt{3}}:\quad\widehat{H^{\beta}[\varepsilon N]}\Psi:=
             \frac{6}{\ellp^4}\Big[\big[\widehat{\boldsymbol{V}}[N],\widehat{H_{\mathrm{E}}}[1]\big],\widehat{H_{\mathrm{E}}}[1]\Big]\Psi
             \label{Qpropsd}
\end{equation}
A moment of reflection reveals, that in the passage from \eref{Hpropsd} to 
the proposed operator version in \eref{Qpropsd}, there appear to be almost 
no quantization ambigiuities. The quantization of the Poisson bracket is 
straightforward, and all remaining quantization ambiguities arise from 
$\widehat{\boldsymbol{V}}[N]$, $\widehat{H_{\mathrm{E}}}[1]$ 
and of course $G_i[e^{ia}\partial_a N]$ respectively. In classical GR 
the functional $G_i[e^{ia}\partial_a N]$ does not introduce any new 
constraints, and can always be absorbed into a redefinition of the 
Lagrange multiplier field $\Lambda^i$ , which we will do in the following. 
Using a so-called internal regularisation for the volume operator as in \cite{ashvolume, status} and provided that $\widehat{H_{\mathrm{E}}}[N]$ is anomaly free \emph{in the sense of Thiemann} the same holds for the proposed quantization in \eref{Qpropsd} too. This immediately follows from the fact that $\widehat{\boldsymbol{V}}[N]$, $\widehat{H_{\mathrm{E}}}[N]$ as well as $\widehat{H_{\mathrm{E}}}[1]$ are not acting on newly created vertices.

At the end of this paper recognising the importance of \cite{parametercorichi, parametermeissner} we have to face the crucial question of the physical relevance of our observation. Do we have to accept the proposed value \eref{bivalue} found from black hole entropy calculations? Arguing that a semi-classical limit of LQG leads to GR for this particular value only this question is commonly answered in the affirmative (see for instance \cite{status}). To our knowledge there is however no indication other than the famous black hole entropy calculations for this assumption available. Furthermore GR is consistent with any value of the Barbero--Immirzi parameter and the result of Bekenstein and Hawking was found in a semi-classical (low energy) approach. From this observation Jacobson proposes in \cite{parameterjacobson} to distinguish between a microscopic entropy $S_{\mathrm{LQG}}$ found directly from LQG and the macroscopic result $S_{\mathrm{BH}}$. He then suggests that microscopic \qq{Planckian} quantities and their low energy counterparts are related to one another by unknown dimensionless functions of $\beta$, and concludes that demanding $S_{\mathrm{LQG}}=S_{\mathrm{BH}}(\beta)$ could be used to fix the Barbero--Immirzi parameter to possibly other values than those found already in the literature \cite{parametercorichi, parametermeissner}.

Let us remark that even if the above value for $\beta$ is not chosen to be $1/\sqrt{3}$ it is still possible to rewrite the Lorentzian constraint as double-bracket upt to an additional additive Euclidean constraint. 
After some algebraic manipulations and again neglecting all Poisson brackets involving $\varepsilon$ (remember that we always set $\{\varepsilon,\cdot\}=0$) one arrives at the following expression:
\begin{equation}
\eqalign{H^{\beta}[N]=\frac{3\beta^2-1}{4\beta^2}H_{\mathrm{E}}[N]
         & -\frac{1+\beta^2}{2\beta^2}\frac{1}{\gamma^2}\big\{\boldsymbol{V}[\varepsilon N],H_{\mathrm{E}}[1]\big\}^2+\\
         & +\frac{1+\beta^2}{4\beta^2}G_i\big[\varepsilon e^{ia}\partial_aN\big]}
\end{equation}
In order to construct a consistent quantization of the above equation it would be needed to either implement the oriented volume of space $\boldsymbol{V}[\varepsilon N]=\int_\Sigma N \omega_E$ quantum theoretically, or to restrict ourselves to the subspace of $\mathcal{K}$ corresponding to fixed orientation $\varepsilon=\pm1$.
\section*{Acknowledgments}
The authors would like to thank Peter C. Aichelburg and Florian Preis for many inspiring discussions.

\end{document}